\documentclass[fleqn,10pt]{wlpeerj}
\usepackage{url}

\title{Quantifying the Effect of Sentiment on Information Diffusion in Social Media}

\author[1,2]{Emilio Ferrara}
\author[1]{Zayao Yang}
\affil[1]{School of Informatics and Computing, Indiana University, Bloomington, IN, USA}
\affil[2]{Indiana University Network Science Institute, Bloomington, IN, USA}


\begin{abstract}
Social media have become the main vehicle of information production and consumption online. 
Millions of users every day log on their Facebook or Twitter accounts to get updates and news, read about their topics of interest, and become exposed to new opportunities and interactions.
Although recent studies suggest that the contents users produce will affect the emotions of their readers, we still lack a rigorous understanding of the role and effects of contents sentiment on the dynamics of information diffusion. 
This work aims at quantifying the effect of sentiment on information diffusion, to understand: \emph{(i)} whether positive conversations spread faster and/or broader than negative ones (or vice-versa); \emph{(ii)} what kind of emotions are more typical of popular conversations on social media; and, \emph{(iii)} what type of sentiment is expressed in conversations characterized by different temporal dynamics. Our findings show that, at the level of contents, negative messages spread faster than positive ones, but positive ones reach larger audiences, suggesting that people are more inclined to share and favorite positive contents, the so-called \emph{positive bias}. As for the entire conversations, we  highlight how different temporal dynamics exhibit different sentiment patterns: for example, positive sentiment builds up for highly-anticipated events, while unexpected events are mainly characterized by negative sentiment. 
Our contribution is a milestone to understand how the emotions expressed in short texts affect their spreading in online social ecosystems, and may help to craft effective policies and strategies for content generation and diffusion.
\end{abstract}

\begin{document}

\flushbottom
\maketitle
\thispagestyle{empty}

\section*{Introduction}
The emerging field of \emph{computational social science} has been focusing on studying the characteristics of techno-social systems \cite{lazer2009life, vespignani2009predicting, kaplan2010users, asur2010predicting, cheng2014can} to understand the effects of technologically-mediated communication on our society  \cite{gilbert2009predicting, ferrara2012large, tang2012inferring, demeo2014facebook, backstrom2014romantic}.
Research on information diffusion focused on the complex dynamics that characterize social media discussions \cite{java2007we, huberman2008social, bakshy2012role, ferrara2013clustering} to understand their role as central fora to debate social issues  \cite{conover2013digital, conover2013geospatial, varol2014evolution}, 
to leverage their ability to enhance situational, social, and political awareness \cite{sakaki2010earthquake, centola2010spread, centola2011experimental, bond201261, ratkiewicz2011detecting, metaxas2012social, ferrara2014rise}, 
or to study susceptibility to influence and social contagion  \cite{aral2009distinguishing, aral2012identifying, myers2012information, anderson2012effects, lerman2010information, ugander2012structural, weng2013virality, weng2014predicting}
The amount of information that generated and shared through online platforms like Facebook and Twitter yields unprecedented opportunities to millions of individuals every day \cite{kwak2010twitter, gomez2013structure, ferrara2013traveling}.
Yet, how understanding of the role of the sentiment and emotions conveyed through the content produced and consumed on these platforms is shallow. 

In this work we are concerned in particular with quantifying the effect of sentiment on information diffusion in social networks. Although recent studies suggest that emotions are passed via online interactions  \cite{harris2007investigation, mei2007topic, golder2011diurnal, de2012not, kramer2014experimental}, and that many characteristics of the content may affect information diffusion (e.g., language-related features \cite{nagarajan2010qualitative}, hashtag inclusion \cite{suh2010want}, network structure \cite{recuero2011does}, user metadata \cite{ferrara2014rise}), little work has been devoted to quantifying the extent to which sentiment drives information diffusion in online social media.
Some studies suggested that content conveying positive emotions could acquire more attention \cite{kissler2007buzzwords, bayer2012font, stieglitz2013emotions} and trigger higher levels of arousal \cite{berger2011arousal}, which can further affect feedback and reciprocity \cite{dang2012impact} and social sharing behavior \cite{berger2012makes}. 

In this study, we take Twitter as scenario, and we explore the complex dynamics intertwining sentiment and information diffusion. 
We start by focusing on contents spreading, exploring what effects sentiment has on the diffusion speed and on contents popularity.
We then shift our attention to entire conversations, categorizing them into different classes depending on their temporal evolution: we highlight how different types of discussion dynamics exhibit different types of sentiment evolution.
Our study timely furthers our understanding of the intricate dynamics intertwining information diffusion and emotions on social media.

% You may title this section "Methods" or "Analysis". 
\section*{Materials and Methods}

%%%%%%%%%%%%%%%%%%%%%%%%%%%%%%%%
\subsection*{Sentiment Analysis}
Sentiment analysis was proven an effective tool to analyze social media streams, especially for predictive purposes \cite{pang2008opinion, bollen2011twitter, bollen2011modeling, le2015predictability}.
A number of sentiment analysis methods have been proposed to date to capture contents' sentiment, and some have been specifically designed for short, informal texts \cite{akkaya2009subjectivity, paltoglou2010study, hutto2014vader}.
To attach a sentiment score to the tweets in our dataset,  we here adopt a SentiStrength, a promising sentiment analysis algorithm that, if compared to other tools, provides several advantages: first, it is optimized to annotate short, informal texts, like tweets, that contain abbreviations, slang, and the like. SentiStrength also employs additional linguistic rules for negations, amplifications, booster words, emoticons, spelling corrections, etc.
Research applications of SentiStrength to MySpace data found it particularly effective at capturing positive and negative emotions with, respectively, 60.6\% and 72.8\% accuracy  \cite{thelwall2010sentiment, thelwall2011sentiment, stieglitz2013emotions}.

The algorithm assigns to each tweet $t$ a positive $S^+(t)$ and negative $S^-(t)$ sentiment score, both ranging between 1 (neutral) and 5 (strongly positive/negative). 
Starting from the sentiment scores, we capture the \emph{polarity} of each tweet $t$ with one single measure, the \emph{polarity score} $S(t)$, defined as the difference between positive and negative sentiment scores:

\begin{equation}
	S(t) = S^+(t) - S^-(t).
	\label{eq:polarity}
\end{equation}

The above-defined score ranges between -4 and +4. The former score indicates an extremely negative tweet, and occurs when $S^+(t)=1$ and $S^-(t)=5$. Vice-versa, the latter identifies an extremely positive tweet labeled with $S^+(t)=5$ and $S^-(t)=1$.
In the case $S^+(t)=S^-(t)$ ---positive and negative sentiment scores for a tweet $t$ are the same--- the polarity $S(t)=0$ of tweet $t$ is considered as neutral.

\begin{figure}[t] \centering
	\includegraphics[width=\columnwidth]{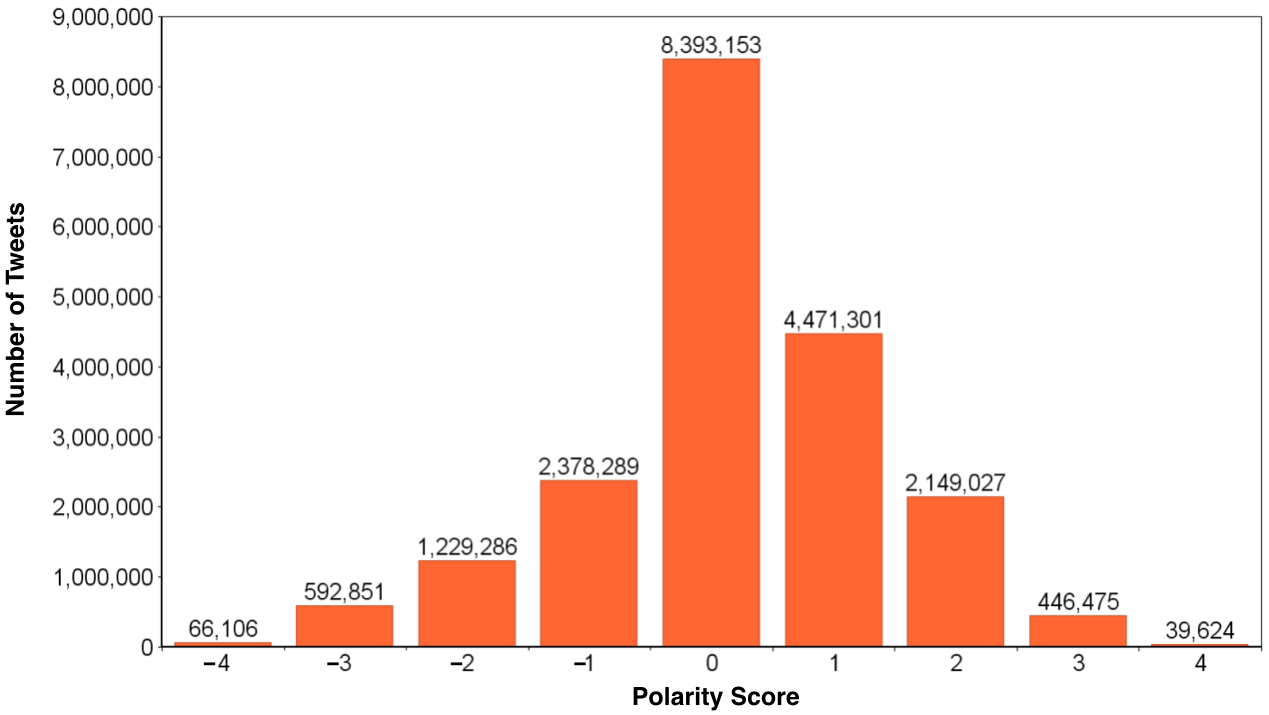} 
	\caption{{\bf Distribution of polarity scores computed for our dataset.} 
	The polarity score $S$ is the difference between positive and negative sentiment scores as calculated by SentiStrength. The dataset ($N=19,766,112$ tweets, by $M=8,130,481$ different users) contains $42.46\%$ of neutral ($S=0$), $35.95\%$ of positive ($S\geq1$), and $21.59\%$ of negative ($S\leq-1$) tweets, respectively.}
	\label{fig:sentiment_distribution}
\end{figure}

%%%%%%%%%%%%%%%%%%
\subsection*{Data}
The dataset adopted in this study contains a sample of all public tweets produced during September 2014. From the Twitter \emph{gardenhose} (a roughly 10\% sample of the social stream that we process and store at Indiana University) we extracted all tweets in English that do not contain URLs or media content (photos, videos, etc.) produced in that month. 
This choice is dictated by the fact that we can hardly computationally capture the sentiment or emotions conveyed by multimedia content, and processing content from external resources (such as webpages, etc.) would be computationally hard. 
This dataset comprises of 19,766,112 tweets (more than six times larger than the Facebook experiment \cite{kramer2014experimental}) produced by 8,130,481 distinct users. 
All tweets are processed by SentiStrength and attached with sentiment scores (positive and negative) and with the polarity score calculated as described before. 
We identify three classes of tweets' sentiment: negative (polarity score $S\leq-1$), neutral ($S=0$), and positive ($S\geq1$). 
Negative, neutral, and positive tweets account for, respectively, 21.59\%, 42.46\% and 35.95\% of the total. 
The distribution of polarity scores is captured by Fig. \ref{eq:polarity}: we can see it is peaked around neutral tweets, accounting for over two-fifth of the total, while overall the distribution is slightly skewed toward positiveness.
We can also observe that extreme values of positive and negative tweets are comparably represented: for example, there are slightly above 446 thousand tweets with polarity score $S=+3$, and about 592 thousands with opposite polarity of $S=-3$.

% Results and Discussion can be combined.
\section*{Results}
\begin{figure*}[!t] \centering
	\includegraphics[width=\columnwidth,clip=true,trim=170 0 170 0 ]{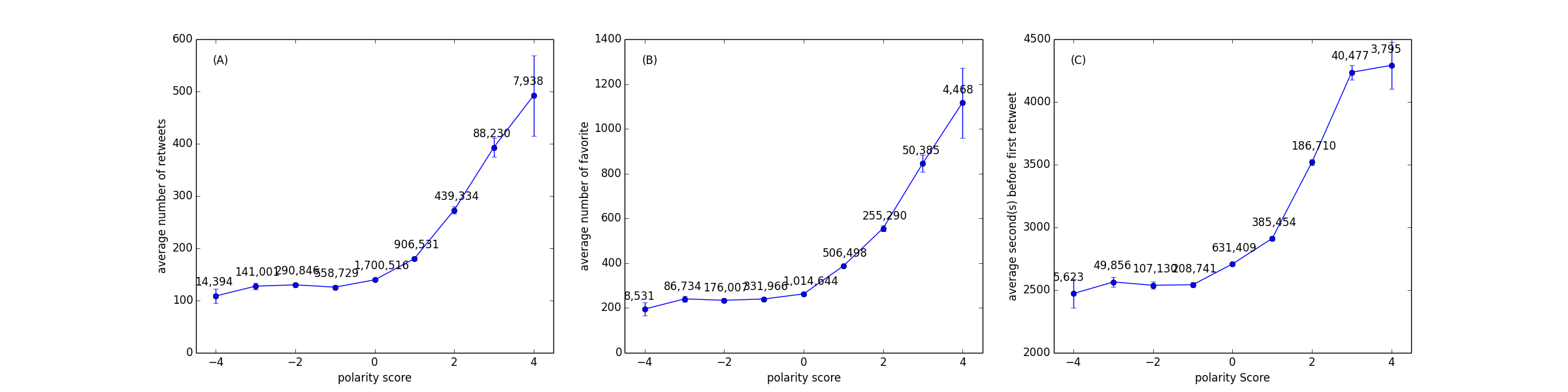}
	\caption{{\bf The effect of sentiment on information diffusion} The three panels show, respectively, (A) the average number of retweets, (B) the average number of favorites, and (C) the average number of seconds passed before the first retweet, as a function of the polarity score of the given tweet. The number on the points represent the amount of tweets with such polarity score in our sample. Bars represent standard errors.}
	\label{fig:attention_classes}
\end{figure*}

%%%%%%%%%%%%%%%%%%%%%%%%%%%%%%%%%%%%%%%%%%%%%%%%%%%%%%%%%%%
\subsection*{The role of sentiment on information diffusion}
Here we are concerned with studying the relation between content sentiment and information diffusion.
Fig. \ref{fig:attention_classes} shows the effect of contents sentiment on the information diffusion dynamics and on contents popularity. 
We measure three aspects of information diffusion, as function of tweets polarity scores: Fig. \ref{fig:attention_classes}A shows the average number of retweets collected by the original posts as function of the polarity expressed therein; 
similarly, Fig. \ref{fig:attention_classes}B shows the average number of times the original tweet has been favored;
Fig. \ref{fig:attention_classes}C illustrates the speed of information diffusion, as reflected by the average number of seconds that occur between the original tweet and the first retweet. 
Both Fig. \ref{fig:attention_classes}A and Fig. \ref{fig:attention_classes}C focus only on tweets that have been retweeted at least once. Fig. \ref{fig:attention_classes}B considers only tweets that have been favored at least once. Note that, a large fraction of tweets never become retweeted (79.01\% in our dataset) or favored (87.68\%): Fig. \ref{fig:attention_classes}A is based on the 4,147,519 tweets that have been retweeted at least once ($RT\geq1$), Fig. \ref{fig:attention_classes}B reports on the 2,434,523 tweets that have favored at least once, and Fig. \ref{fig:attention_classes}C comprises of the 1,619,195 tweets for which we have observed the first retweet in our dataset (so that we can compute the time between the original tweet and the first retweet).
Note that the retweet count is extracted from the tweet metadata, instead of being calculated as the number of times we observe a retweet of each tweet in our dataset, in order to avoid the bias due to the sampling rate of the Twitter \emph{gardenhose}. 
For this reason, the average number of retweets reported in Fig. \ref{fig:attention_classes}A seems pretty high (above 100 for all classes of polarity scores): by capturing the ``true'' number of retweets we well reflect the known broad distributions of content popularity of social media, skewing the values of the means toward larger figures.
The very same reasoning applies for the number of favorites.

Two important considerations emerge from the analysis of Fig. \ref{fig:attention_classes}: \emph{(i)} positive tweets spread broader than neutral ones, and collect more favorites, but interestingly negative posts do not spread any more or less than neutral ones, neither get more or less favored. 
This suggests the hypothesis of observing the presence of \emph{positivity bias} \cite{garcia2012positive} (or \emph{Pollyanna hypothesis} \cite{boucher1969pollyanna}), that is the tendency of individuals to favor positive rather than neutral or negative items, and choose what information to favor or rebroadcast further accordingly to this bias. 
\emph{(ii)} Negative contents spread much faster than positive ones, albeit not significantly faster than neutral ones. 
This suggests that positive tweets require more time to be rebroadcasted, while negative or neutral posts generally achieve their first retweet twice as fast. 
Interestingly, previous studies on information cascades showed that all retweets after the first take increasingly less time, which means that popular contents benefit from a feedback loop that speeds up the diffusion more and more as a consequence of the increasing popularity \cite{kwak2010twitter}.

%%%%%%%%%%%%%%%%%%%%%%%%%%%%%%%%%%%%%%%%%%%%%%%%%%%%%%%%%%%%%
\subsection*{Conversations' dynamics and sentiment evolution}
\begin{figure}[!t] \centering
	\includegraphics[width=\columnwidth,clip=true,trim=0 0 45 0]{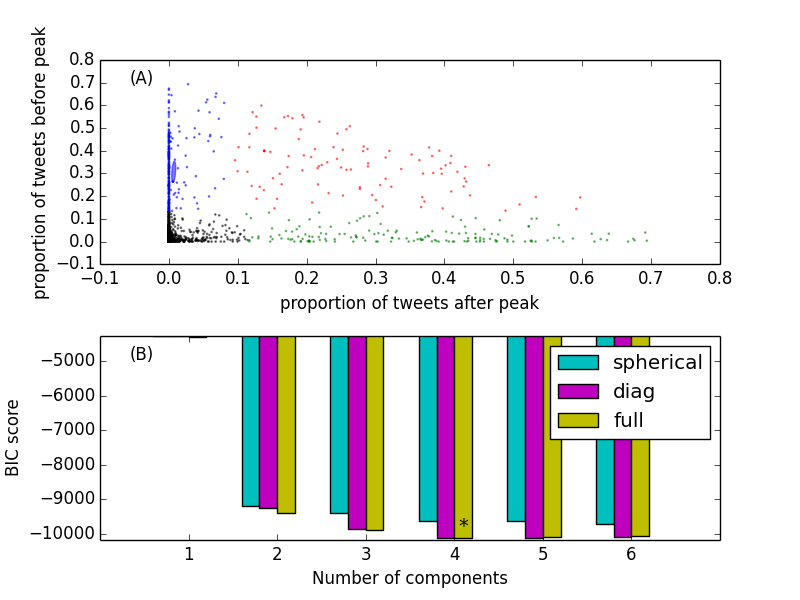} 
	\caption{{\bf Dynamical classes of popularity capturing four different types of Twitter conversations.}
	Panel (A) shows the Gaussian Mixture Model employed to discover the four classes.
	 The y and x axes represent, respectively, the proportion of tweets occurring before and after the peak of popularity of a given discussion. Different colors represent different classes: anticipatory discussions (blue dots), unexpected events (green), symmetric discussions (red), transient events (black).
	 Panel (B) shows the BIC scores of different number of mixture components for the GMM (the lower the BIC the better the GMM captures the data). The star identifies the optimal number of mixtures, four, best captured by the full model.}
	\label{fig:hashtag_classification}
\end{figure}

To investigate how sentiment affects content popularity, we now only consider active and exclusive discussions occurred on Twitter in September 2014. Each topic of discussion is here identified by its most common hashtag. Active discussions are defined as those with more than 200 tweets (in our dataset, which is roughly a 10\% sample of the public tweets), and exclusive ones are defined as those whose hashtag never appeared in the previous (August 2014) and the next (October 2014) month.

Inspired by previous studies that aimed at finding how many types of different conversations occur on Twitter \cite{kwak2010twitter, lehmann2012dynamical}, we characterize our discussions according to three features: the proportion $p_b$ of tweets produced within the conversation before its peak, the proportion $p_d$ of tweets produced during the peak, and finally the proportion $p_a$ of tweets produced after the peak. 
The peak of popularity of the conversation is simply the day which exhibits the maximum number of tweets with that given hashtag.
We use the Expectation Maximization (EM) algorithm to learn an optimal Gaussian Mixture Model (GMM) in the $(p_b, p_a)$ space.
To determine the appropriate number of components (i.e., the number of types of conversations), we adopt three GMM models (spherical, diagonal, and full) and perform a 5-fold cross-validation using the Bayesian Information Criterion (BIC) as quality measure. We vary the number of components from 1 to 6. 
Fig. \ref{fig:hashtag_classification}B  shows the BIC scores for different number of mixtures: the lower the BIC score, the better. 
The outcome of this process determines that the optimal number of components is four, in agreement with previous studies \cite{lehmann2012dynamical}, as captured the best by the full GMM model.
In Fig. \ref{fig:hashtag_classification}A we show the optimal GMM that identifies the four classes of conversation: the two dimensions represent the proportion $p_b$ of tweets occurring before (y axis) and $p_a$ after (x axis) the peak of popularity of each conversation.

\begin{figure}[!t] \centering
	\includegraphics[width=\columnwidth]{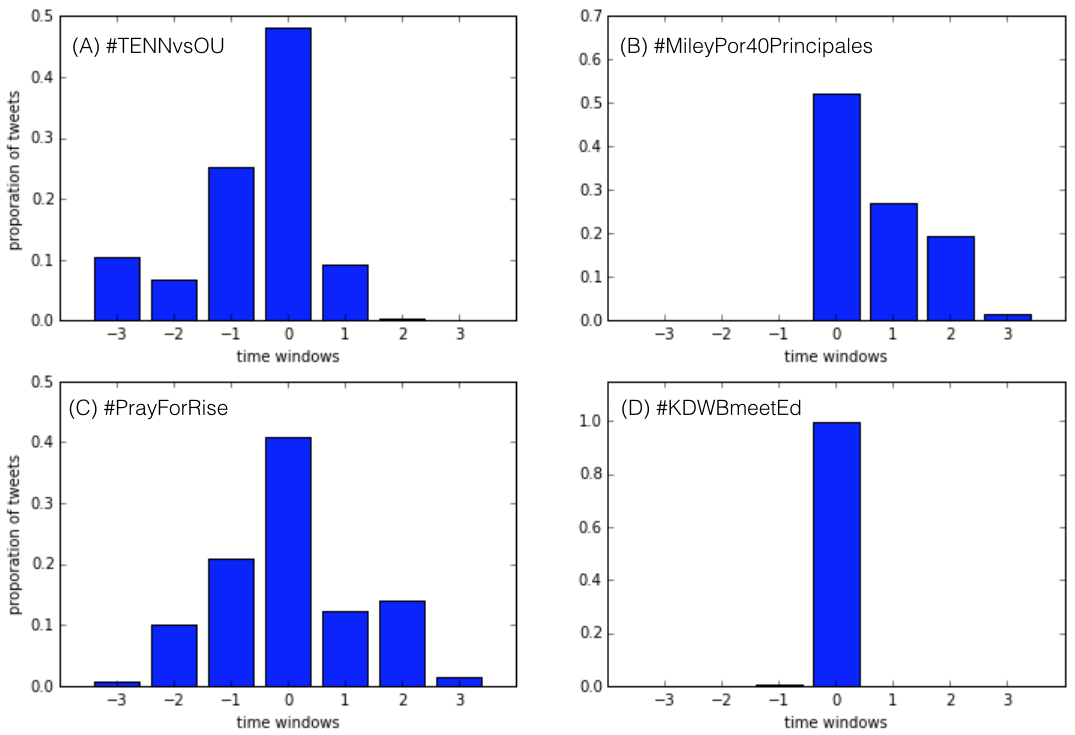} 
	\caption{{\bf Example of four types of Twitter conversations reflecting the respective dynamical classes, in our dataset.} Panel (A) shows one example of anticipatory discussion (\#TENNvsOU); (B) an unexpected event (\#MileyPor40Principales); (C) a symmetric discussion (\#PrayForRise); and, (C) a transient event (\#KDWBmeetEd).}
	\label{fig:typical_distribution}
\end{figure}

The four classes correspond to: \emph{(i) anticipatory discussions} (blue dots), \emph{(ii) unexpected events} (green), \emph{(iii) symmetric discussions} (red), and \emph{(iv) transient events} (black).
Anticipatory conversations (blue) exhibit most of the activity before and during the peak. These discussions build up over time registering an anticipatory behavior of the audience, and quickly fade out after the peak.
The complementary behavior is exhibited by discussions around unexpected events (green dots): the peak is reached suddenly as a reaction to some  exogenous event, and the discussion quickly decays afterwards.
Symmetric discussions (red dots) are characterized by a balanced number of tweets produced before, during, and after the peak time. 
Finally, transient discussions (black dots) are typically bursty but short events that gather a lot of attention, yet immediately phase away afterwards.
According to this classification, out of 1,522 active and exclusive conversations (hashtags) observed in September 2014, we obtained 64 hashtags of class A (anticipatory), 156 of class B (unexpected), 56 of class C (symmetric), and 1,246 of class D (transient), respectively.
Fig. \ref{fig:typical_distribution} shows examples representing the four dynamical classes of conversations registered in our dataset.
The conversation lengths are all set to 7 days, and centered at the peak day (time window 0).

Fig. \ref{fig:typical_distribution}A represents an example of anticipatory discussion: the event captured (\#TENNvsOU) is the football game \emph{Tennessee Volunteers vs. Oklahoma Sooners} of Sept. 13, 2014. 
The anticipatory nature of the discussion is captured by the increasing amount of tweets generated before the peak (time window 0) and by the drastic drop afterwards. 
Fig. \ref{fig:typical_distribution}B shows an example (\#MileyPor40Principales) of discussion around an unexpected event, namely the release by \emph{Los 40 Principales} of an exclusive interview to Miley Cyrus, on Sept. 10, 2014. 
There is no activity before the peak point, that is reached immediately the day of the news release, and after that the volume of discussion decreases rapidly.
Fig. \ref{fig:typical_distribution}C represents the discussion of a symmetric event: \#PrayForRise was a hashtag adopted to support RiSe, the singer of the K-pop band \emph{Ladies' Code}, who was involved in a car accident that eventually caused her death. 
The symmetric activity of the discussion perfectly reflects the events:\footnote{Wikipedia: Ladies' Code --- \url{http://en.wikipedia.org/wiki/Ladies\%27_Code}} the discussion starts the day of the accident, on  September 3, 2014, and peaks the day of RiSe's death (after four days from the accident, on September 7, 2014), but the fans' conversation stays alive to commemorate her for several days afterwards.
Lastly, Fig. \ref{fig:typical_distribution}D shows one example (\#KDWBmeetEd) of transient event, namely the radio station KDWB announcing a lottery drawing of the tickets for Ed Sheeran's concert, on Sept. 15, 2014. 
The hype is momentarily and the discussion fades away immediately after the lottery is concluded.

\begin{figure}[!t] \centering
	\includegraphics[width=\columnwidth]{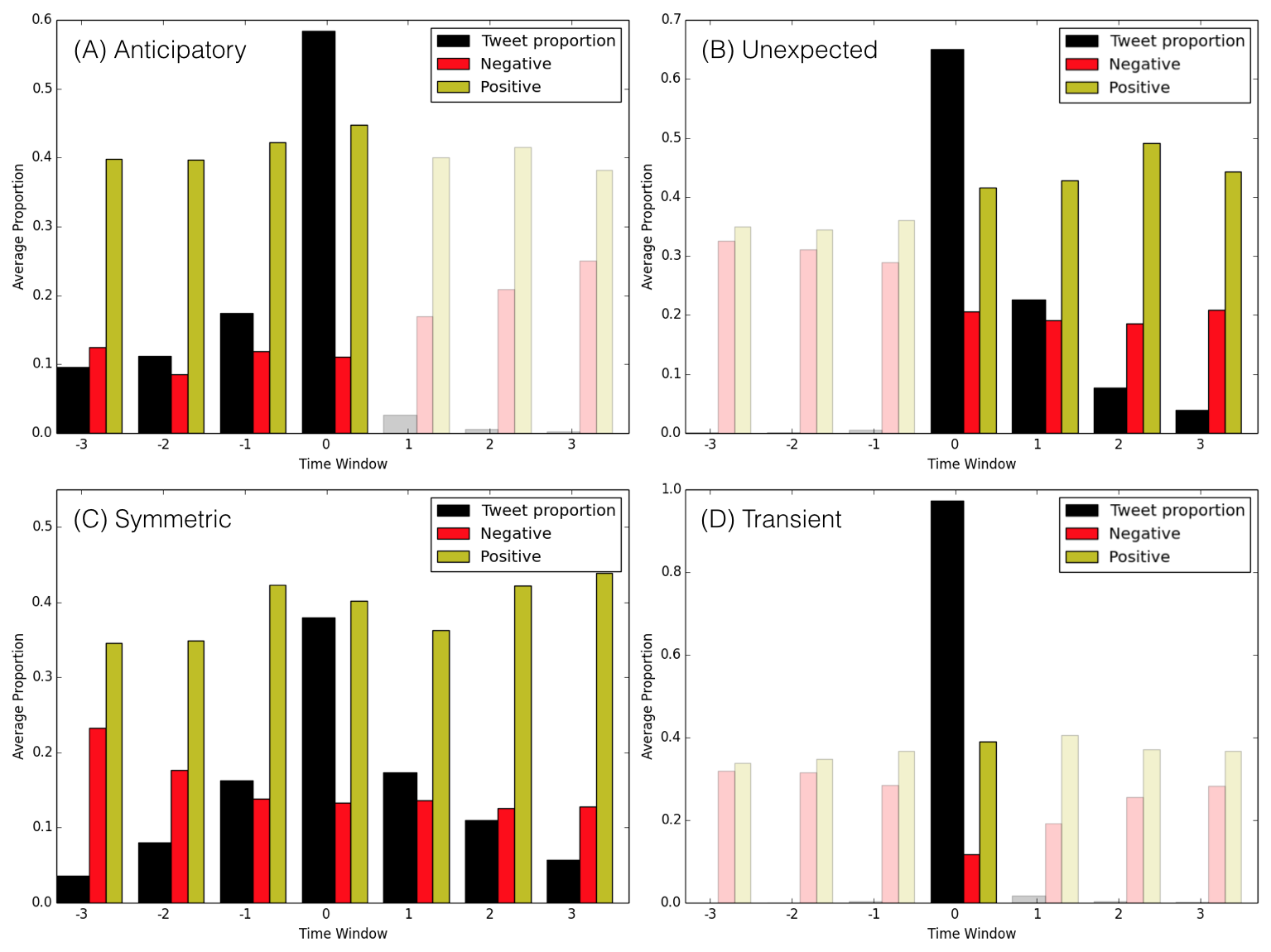} 
	\caption{{\bf Evolution of positive and negative sentiment for different types of Twitter conversations.} The four panels show the average distribution of tweet proportion, and the average positive ($S\geq1$) and negative ($S\leq-1$) tweet proportions, for the four classes respectively: (A) anticipatory discussion; (B) unexpected event; (C) symmetric discussion; and, (D) transient discussion. Grayed bars in A, B, and D, represent non-significant amounts of tweets for non-representative time windows of those classes of conversations.}
	\label{fig:class_evolution}
\end{figure}

Fig. \ref{fig:class_evolution} shows the evolution of sentiment for the four classes of Twitter conversations: it can be useful to remind the average proportions of neutral ($42.46\%$), positive ($35.95\%$), and negative ($21.59\%$) sentiments in our dataset, to compare them against the distributions for popular discussions.
\emph{(A)} For anticipatory events, the amount of positive sentiment grows steadily until the peak time, while the negative sentiment is somewhat constant throughout the entire anticipatory phase. Notably, the amount of negative content is much below the dataset average, fluctuating between 9\% and 12\% (almost half of the dataset average), while the positive content is well above average, ranging between 40\% and 44\%. This suggests that, in general, anticipatory popular conversations are emotionally positive. 
(B) The class of unexpected events intuitively carries more negative sentiment, that stays constant throughout the entire discussion period to levels of the dataset average. 
(C) Symmetric popular discussions are characterized by a steadily decreasing negative emotions, that goes from about 23\% (above dataset's average) at the inception of the discussions, to around 12\% toward the end of the conversations. Complementary behavior happens for positive emotions, that start around 35\% (equal to the dataset average) and steadily grow up to 45\% toward the end. This suggests that, in symmetric conversations there is a general shift of emotions toward positiveness over time. 
(D) Finally, transient events, due to their short-lived lengths, represent more the average discussions, although they exhibit lower levels of negative sentiments (around 15\%) and higher levels of positive ones (around 40\%) with respect to the dataset's averages.

\section*{Discussion}
The ability to computationally annotate at scale the emotional value of short pieces of text, like tweets, allowed us to investigate the role that emotions and sentiment expressed into social media contents play with respect to the diffusion of such information. 

Our first finding in this study sheds light on how sentiment affects the speed and the reach of the diffusion process: tweets with negative emotional valence spread faster than neutral and positive ones. In particular, the time that passes between the publication of the original post and the first retweet is almost twice as much, on average, for positive tweets than for negative ones. This might be interpreted in a number of ways, the most likely being that contents that convey negative sentiments trigger stronger reactions in the readers, some of which might be more prone to reshare that piece of information with higher chance than any neutral or positive content. However, the \emph{positive bias} (or Pollyanna effect) \cite{garcia2012positive, boucher1969pollyanna} rapidly kicks in when we analyze how many times the tweets become retweeted or favored: individuals online clearly tend to prefer positive tweets, which are favored as much as five times more than negative or neutral ones; the same holds true for the amount of retweets collected by positive posts, which is up to 2.5 times more than negative or neutral ones. These insights provide some clear directives in terms of best practices to produce popular content: if one aims at triggering a quick reaction, negative sentiments outperform neutral or positive emotions. This is the reason why, for example, in cases of emergencies and disasters, misinformation and fear spread so fast in online environments \cite{ferrara2014rise}.
However, if one aims at long-lasting diffusion, then positive contents ensure wide reach and the most preferences. 

The second part of our study focuses on entire conversations, and investigates how different sentiment patterns emerge from discussions characterized by different temporal signatures~\cite{kwak2010twitter, lehmann2012dynamical}: we discover that, in general, highly-anticipated events are characterized by positive sentiment, while unexpected events are often harbinger of negative emotions; yet, transient events, whose duration is very brief, represent the norm on social media like Twitter and are not characterized by any particular emotional valence. These results might sound unsurprising, yet they have not been observed before: common sense would suggest, for example, that unprecedented conversations often relate to unexpected events, such as disasters, emergencies, etc., that canalize vast negative emotions from the audience, including fear, sorrow, grief, etc.~\cite{sakaki2010earthquake} Anticipated conversations instead characterize events that will occur in the foreseeable future, such as a political election, a sport match, a movie release, an entertainment event, or a recurring festivity: such events are generally positively received, yet the attention toward them quickly phases out after their happening~\cite{lehmann2012dynamical, mestyan2013early, le2015predictability}. Elections and sport events might represent special cases, as they might open up room for debate, ``flames'', polarized opinions, etc.~\cite{ratkiewicz2011detecting, bond201261} (such characteristics have indeed been exploited to make predictions~\cite{asur2010predicting, metaxas2012social, le2015predictability}).

The findings of this paper have very practical consequences, that are relevant both for economic and social impact: understanding the dynamics of information diffusion and the effect of sentiment on such phenomena becomes crucial if one, for example, wants to craft a policy to effectively communicate with an audience. The applications range from advertisement and marketing, to public policy and emergency management. Recent events, going for tragic episodes of terrorism, to the emergence of pandemics like Ebola, have highlighted once again how central social media are in the timely diffusion of information, yet how dangerous they can be when they are abused or misused to spread misinformation or fear. Our contribution sets a first milestone to understand how the emotions expressed in a short piece of text might affect its spreading in online social ecosystems, helping to craft effective information diffusion strategies that account for the emotional valence of the contents.

\bibliographystyle{abbrv}
\bibliography{references}

\end{document}